\documentclass[prl,amsmath,amssymb,reprint, superscriptaddress,nofootinbib,floatfix]
{revtex4-1}
\usepackage{amsmath,amssymb}
\usepackage{hyperref}
\usepackage{graphicx}

\usepackage{tikz}

\begin{document}
\title{Rigorous construction of coherent state path integrals through dualization}
\newcommand{\RegensburgUniversity}{Institut f\"ur Theoretische Physik, 
Universit\"at Regensburg, D-93040 Regensburg, Germany}
\author{Falk Bruckmann}
\affiliation{\RegensburgUniversity}
\author{Juan Diego Urbina}
\affiliation{\RegensburgUniversity}

\begin{abstract}
Two long-standing problems in the construction of coherent state path integrals, the unwarranted assumption of path continuity and the ambiguous definition of the Hamiltonian symbol, are rigorously solved. To this end the fully controlled dual representation familiar from lattice quantum field theories is introduced. Dualization allows for both the step-by-step check in the construction of discrete path integrals and for the identification of the Hamiltonian and Berry phase part of the action, thus making {\it both} discrete and continuous path integrals consistently defined. Once the correct action is constructed, we are able to follow the transition to the continuum for the polar form of general Bose-Hubbard models and to provide an exact form of the path integral for general spin systems, where previous works showed its failure for all standard choices of operator ordering.   
\end{abstract}
%
\maketitle

Seventy years after its appearance in quantum mechanics, Feynman's path integral with its elegance and intuitive appeal is an essential part of the theoretical Physics toolbox. In fact, many quantum (field) theories are defined by just a path integral that provides their -- nonperturbative -- quantization. However, the persistent issue of the transition from the discrete to the (formal) continuous versions of the theory remains. Important physical phenomena like topological effects and key approximation methods like semiclassical expansions both rely on the continuum versions, while for their numerical evaluation path integrals are typically re-discretized (also controling their infinities) in an ad hoc manner. 

\begin{figure*}[!t]
  \begin{tikzpicture}
  \node (Hcorr) at (-8.0,1.2) 
    [rounded corners=3pt, thick, draw=green, minimum height=2.0em]
    {$\sum_k\left[z_{k}^{*}(z_{k}-z_{k+1})
    +\Delta H(z_{k}^{*},z_{k+1})\right]
    \quad\eqref{Eq.rus}
    $};
  \node (Hdiag) at (-6.7,0) 
    [rounded corners=3pt, thick, draw=red, minimum height=2.0em]
    {$\sum_k\left[z_{k}^{*}(z_{k}-z_{k+1})
    +\Delta H(z_{k}^{*},z_{k})\right]\quad(\text{I})$};
  \node (Hcont) at (-8.8,-1.2) 
    [rounded corners=3pt, thick, draw=red, minimum height=2.0em] 
    {$\int\!d\tau\left[-
    i\rho\,\dot{\varphi}
    +H(\rho;\varphi)\right]\quad(\text{II})$};
  \draw [<-] (-9.9,-0.8) -- (-9.9,0.8);
  \node (GW) at (-10.3,0) {WG};
  \draw [<-] (-7.5,0.4) -- (-7.5,0.8);
  \node (eins) at (-8.4,0.6) {below \eqref{Eq.aus}};
  \draw [->] (-7.5,-0.4) -- (-7.5,-0.8) [dashed];
  \node (hcorr) at (-0.5,0) 
    [rounded corners=3pt, thick, draw=green, minimum height=2.0em]
    {$\sum_k\left[z_{k}^{*}(z_{k}-z_{k+1})
    +\Delta h(z_{k}^{*},z_{k})\right]
    \quad\eqref{Eq.uru}
    $};
  \node (hcont) at (-0.1-1.1,-1.2) 
    [rounded corners=3pt, thick, draw=red, minimum height=2.0em] 
    {$\int\!d\tau\left[-
    i\rho\,\dot{\varphi}
    +h(\rho;\varphi)\right]\quad(\text{III})$};
  \draw [->] (-0.1-1.1,-0.4) -- (-0.1-1.1,-0.8);
  \node (zwei) at (-1.1-0.9,-0.6) {a la WG};

   \node (uscont) at (4.2,-1.2) 
    [rounded corners=3pt, thick, draw=green, minimum height=2.05em] 
    {$\int\!d\tau\big[-i\rho\,\dot{\phi}
    +{\cal H}(\rho;\phi)\big]$
    \quad \eqref{Eq.cro}-\eqref{Eq.nig2}};
   \node (dualv) at (5.1,-0.5) [] {dual variables};
   \node (aux1) at (2.3,0.0) {};
   \node (aux2) at (4.0,-0.9) {};
   \draw [->] (aux1) to [out=0, in =90] (aux2);
   \node (ham) at (-0.3,1.2) [] {$\hat{H}$};
  \draw [double,->] (ham) -- (-4.9,1.2);  
  \draw [double,->] (ham) -- (-0.3,0.4); 
 \end{tikzpicture}
 \caption{Correct (green) and incorrect (red) bosonic actions for coherent state path integrals with symbols discussed in the text.}
\label{Fig.wrong}
\end{figure*}
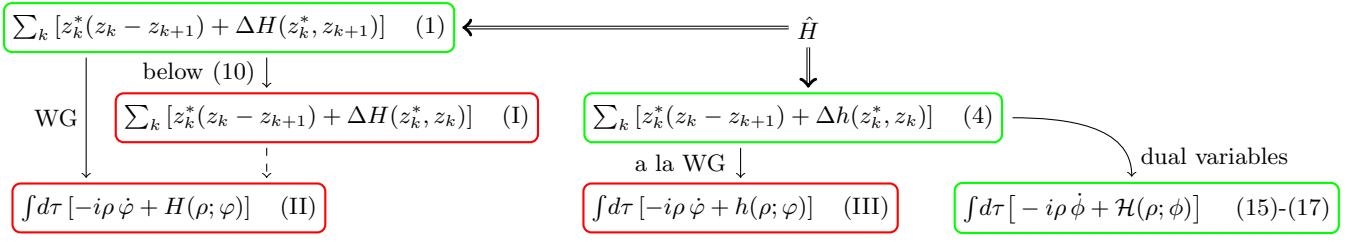

Even for the cleanest example, i.e.\ nonrelativistic thermodynamic partition functions given as a trace over periodic boundary conditions in Euclidean time, and despite great efforts, the equivalence of discrete path integrals against both the continuum limit and the results from the operator formulation has not been established beyond linear systems, and every textbook spends a section warning of the transition from discrete to continuous paths \cite[Sec.~6]{Klauder}, \cite[Sec.~7.7]{Kleinert}, \cite[Sec.~27]{Schulman}, \cite[Sec~2.6]{Kamenev}, \cite[Sec.~7.2]{Fradkin}, \cite[Sec.~12.3]{Coleman} and the related problems with operator ordering. These issues are known to be particularly problematic for {\it coherent state path integrals} and their widely used polar form \cite{Negele,Altland,Coleman}, the latter unavoidable in the case of spin systems \cite{Schulman,Fradkin,Altland}. Pointing at continuity and ordering problems beyond the harmonic case -- where specially designed regularization schemes exist based on exact Gaussian integration \cite{Kleinert} -- in \cite{Wilson} Wilson and Galitski (WG) showed the breakdown of the polar form of the coherent path integral for the one-site Bose-Hubbard model and single spin systems with Ha\-mil\-to\-nians quadratic in the spin generator: the result takes the form of a partition function, but it does not reproduce the correct one obtained from operator methods.

In a nutshell, for a system with Hamiltonian $\hat{H}$ and an overcomplete coherent state basis labelled by complex numbers $z$ eventually building discrete sequences (paths) $\{z_{1},..,z_{N}\}$, there are different choices for the complex function $H(z^*\!,\!z')$, the symbol, representing the unique $\hat{H}$. This itself is not a problem in an exact treatment, as all representations must then be equivalent. The problems, however, come from two unwarranted approximations. First, ordering and continuity get mixed by the textbook assumption $H(z_k^*,z_{k+1})=H(z_k^*,z_{k})+\mathcal{O}(\Delta)$ when turning to paths $z(\tau)$ defined by $z(k\Delta)=z_k$ with $\Delta\sim 1/N$. Second, discrete actions contain a symplectic term $z_k^*(z_{k+1}-z_{k})$ wrongly transformed to polar coordinates $z=\sqrt{\rho}\,{\rm e}^{i\varphi}$ as $z_k^*(z_{k+1}-z_{k})=i\rho_{k}(\varphi_{k}-\varphi_{k-1})+\mathcal{O}(\Delta)$, via a chain of approximations including $z^*\dot{z}\simeq i\rho\dot{\varphi}$ (under the $\tau$-integral). To hope that the errors of these uncontrolled assumptions compensate each other is a dangerous road, especially because these terms are specific for the system's Hamiltonian vs.\ universal. Fig.~\ref{Fig.wrong} summarizes this highly unsatisfactory situation, where every ordering produces a different symbol (for $h$ see below), leading to different results especially in the ground-state region. 

Our goal here is to construct the coherent state path integral representation of partition functions in general non-linear systems without restoring to any ordering or continuity assumption. In the main part of this letter, we will present a fundamentally new approach where {\it every step of the derivation of the coherent state path integral will be checked} against an exact and controlled expansion in terms of dual variables. We then will proceed to circumvent both ordering and continuity assumptions while respecting the guidance principles of locality and intuitive simplicity that are a hallmark of path integral approaches. It is remarkable that by purely identical transformations, and without {\it any} assumptions about continuity of the paths and/or of the Hamiltonian symbol, this program can be fully (and succesfully) pursued in all the cases presented by WG, while from its very construction it is clear that will remain correct for general Hamiltonians. In a sense, then, what we are providing here is a {\it definition} of the coherent state path integral that can be proven to be strictly equivalent to the operator formulation at every step of its derivation.

For definiteness, we start with a single bosonic mode with Hamilton operator $\hat{H}(\hat{b}^\dagger,\hat{b})$ in terms of creation and annihilation operators with $[\hat{b},\hat{b}^{\dagger}]=\hat{I}$. Coherent states defined by $|z\rangle=\exp(-|z|^2/2+z\,\hat{b}^\dagger)|0\rangle$ with complex $z$ satisfying $b|z\rangle=z|z\rangle$ provide an overcomplete basis and can be used to construct two discrete path integral approximations to the partition function ${\cal Z}={\rm Tr}\, {\rm e}^{-\beta \hat{H}}$ approaching it in the limit $N\to\infty$ (resp. $\beta/N=\Delta\to 0$) \cite{Negele}. We first follow the textbook approach getting \begin{align}
 {\cal Z}^{(N)}_{H}
 &=\int \!dz\,
 {\rm e}^{\,\sum_{k=1}^N\left[z_{k}^{*}(z_{k+1}-z_{k})
 -\Delta H(z_{k}^{*},z_{k+1})\right]} 
\label{Eq.rus} 
\end{align}
with $\int\! dz=\prod_{k=1}^{N}\int d\,\text{Re}\,z_{k}\, d\,\text{Im}\,z_{k}/\pi$ and $z_{N+1}=z_{1}$. The first -- symplectic or Berry phase -- factor comes from the scalar product $\langle z_k|z_{k+1}\rangle$ and the second -- Hamiltonian -- factor contains the {\it in general complex} symbol
\begin{align}
 H(z_{k}^{*},z_{k+1})
 :=\frac{\langle z_{k}|\hat{H}|z_{k+1}\rangle}
 {\langle z_{k}|z_{k+1}\rangle}\,.
 \label{Eq.sar}
\end{align}
It is best computed in normal ordered form ($\hat{b}$ right of $\hat{b}^\dagger$), where one simply replaces $\hat{b}\to z_{k+1}$, $\hat{b}^\dagger\to z_k^*$. A second form is provided by the $P$-representation \cite{Scully},
\begin{align}
 \hat{H}=
 \int \frac{d\,\text{Re}\,z\, d\,\text{Im}\,z}{\pi}\,
 h(z^*,z)\,|z\rangle\langle z|
 \label{Eq.egy}
\end{align}
best computed in anti-normal ordered form ($\hat{b}$ left of $\hat{b}^\dagger$), where $\hat{b}\to z_{k}$, $\hat{b}^\dagger\to z_k^*$.
In contrast to $H$, the {\it real} $h$-symbol is diagonal in arguments $z_k$ in its path integral
\begin{align}
 {\cal Z}^{(N)}_{h}
 &=\int \!dz\,
 {\rm e}^{\,\sum_{k=1}^N \left[z_{k}^{*}(z_{k+1}-z_{k})
 -\Delta h(z_{k}^{*},z_k)\right]}.
\label{Eq.uru} 
\end{align}
 
As the first system we consider {\it nonlinear Hamiltonians} 
\begin{align}
 \hat{H}_q=g\,(\hat{b}^\dagger)^q (\hat{b})^q
\label{Eq.mar}
\end{align}
including the harmonic oscillator for $q=1$, an admittedly special linear case, and the one-mode Bose-Hubbard system discussed in WG and \cite{Kordas}  for $q=2$. The partition function is immediately given in the number operator approach $\hat{n}=\hat{b}^\dagger\hat{b}$ with eigenvalues $n=0,1,..$. One finds 
$(\hat{b}^\dagger)^q (\hat{b})^q
=\hat{n}(\hat{n}-1)..(\hat{n}-q+1)
=:\hat{n}!/(\hat{n}-q)!$  \cite{Supplement}, thus 
\begin{align}
 {\cal Z}_q
 =\sum_{n=0}^{\infty}{\rm e}^{-\beta g \frac{n!}{(n-q)!}}
 =q+\sum_{n=0}^{\infty}{\rm e}^{-\beta g \frac{(n+q)!}{n!}}
\label{Eq.irn}
\end{align}
with a characteristic $q$-fold ground state degeneracy. Since the Hamiltonian in Eq.~\eqref{Eq.mar} is normal ordered, the $H$-symbol is directly obtained as $H_q(z_k^*,z_{k+1})=g(z_k^*)^q(z_{k+1})^q$, and its local approximation reads $H_q(z_k^*,z_{k})=g\,|z_k|^{2q}$. From anti-normal ordering $(\hat{b}^\dagger)^q(\hat{b})^q=\sum_{r=0}^q (-1)^{r+q}\frac{q!}{r!}{q\choose r}\, (\hat{b})^r(\hat{b}^\dagger)^r$ -- which follows from the commutator relation without modifying the Hamiltonian -- one obtains $h_q(z_k^*,z_k)=g\,(-1)^q q!\, L_q(|z_k|^2)$ as a Laguerre polynomial \cite{Supplement}.

\begin{figure}[!b]
  \includegraphics[width=0.95\linewidth]{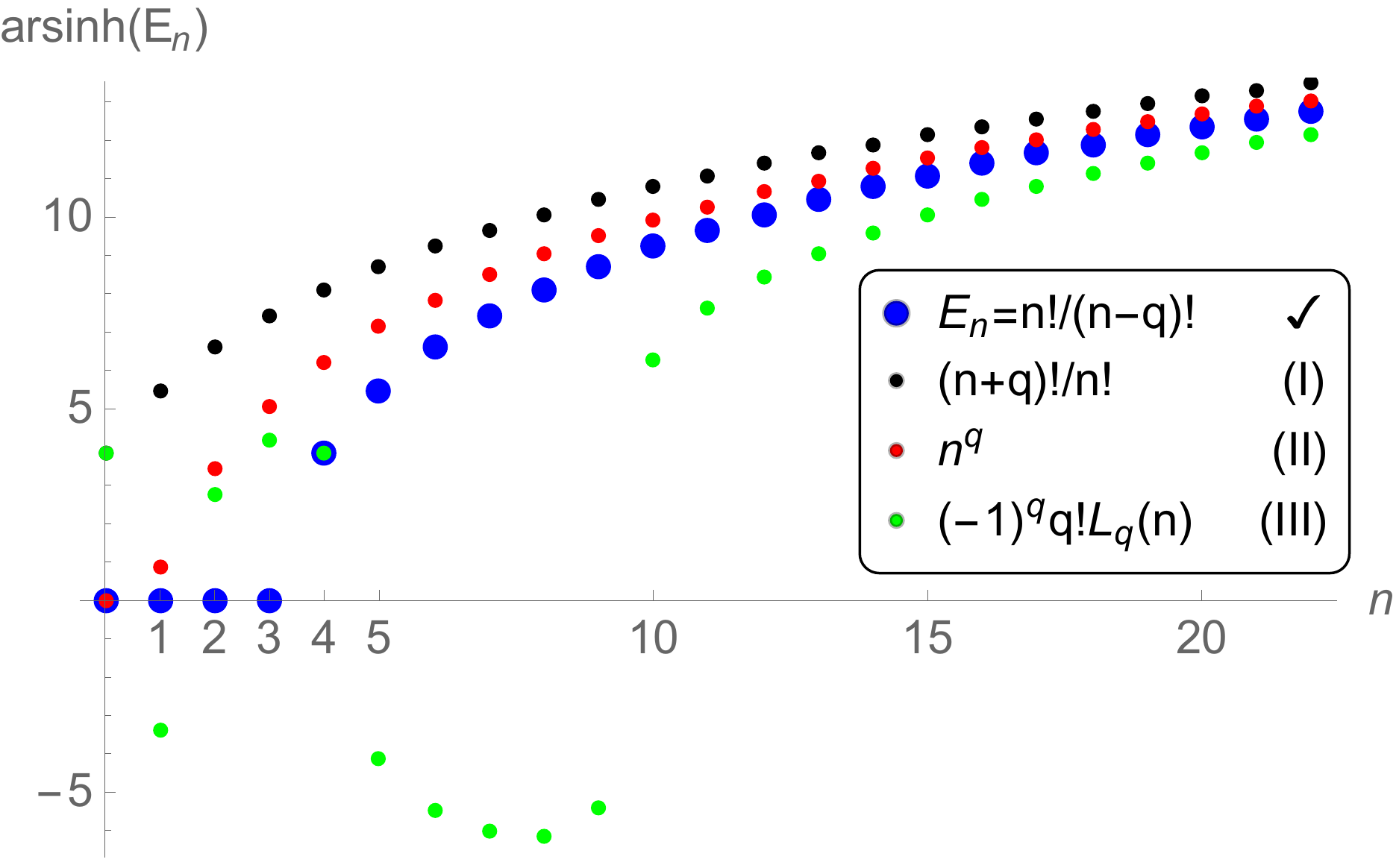}
 \caption{Exponents (squeezed by arcsinh) of the partition function of Eq.~\eqref{Eq.irn} with $q=4$, $\beta g=1$ (blue) and wrong path integral results (I)-(III) for it (black, red, green) discussed in the text and Fig.~\ref{Fig.wrong}. All agree in the asymptotic behavior, but none of the latter gets the four zero ground states correctly.}
 \label{Fig.comparison}
\end{figure}

Before introducing our approach let us summarize and analyse the problem of {\it continuum path integrals} by virtue of this example. While it has been long known that mild problems (extra additive terms) arise in the $q=1$ case \cite{Kleinert}, for $q \ge 2$ the polar form of the continuum action from the $H$-symbol, $\int_0^{\beta}\!d\tau\,(-i\rho\dot{\varphi}+g\rho^{q})$, i.e.\ (II) in Fig.~\ref{Fig.wrong}, was definitively proven wrong by WG: after integration by parts the $\varphi$-integral demands $\rho$ to be constant, while the boundary term fixes it to non-negative integers $n$. Then ${\cal Z}_q=\sum_{n=0}^\infty \exp(-g\beta n^q)$ which differs from the correct result of Eq.~\eqref{Eq.irn} especially in the low-lying states, see Fig.~\ref{Fig.comparison}. The methods of \cite{Kordas} do not apply beyond $q=2$.

Since the continuity of the paths assumed in this result is used at both symplectic (by using standard chain rules for the change to polar coordinates) and Hamiltonian (by using a diagonal symbol) terms of the action, we need a way to disentangle these two issues. To this end the continuity of the symplectic term can be proven {\it independently wrong} by means of the (exactly diagonal) $h$-representation, where a direct calculation in the same fashion, i.e.\ with the action (III) in Fig.~\ref{Fig.wrong}, gives $\sum_{n=0}^\infty \exp(-g\beta (-1)^q q! L_q(n))$, that again differs from the correct result of Eq.~\eqref{Eq.irn}, see Fig.~\ref{Fig.comparison}. 

Our method of tackling path integrals 
consists of {\it dual} (flux/worldline) {\it variables}, an approach that has solved the sign problem at nonzero chemical potential in a number of physical systems \cite{Gattringer,Bruckmann} and also bears similarities to the first step in the dualization of the Ising model \cite{Kramers}. The guiding principle is to factorize the original variables on different time-slices. To this end, the off-diagonal symplectic exponential is expanded as 
\begin{align}
 {\rm e}^{\,z_{k}^{*}z_{k+1}}
 =\sum_{m_{k}=0}^{\infty}\frac{(\rho_{k}\rho_{k+1})^{m_k/2}\,
 {\rm e}^{-i(\varphi_k-\varphi_{k+1})m_k}} {m_{k}!}\,,
\label{Eq.por}
\end{align}
where $(\rho,\varphi)_{N+1}$ $=(\rho,\varphi)_1$, while the measure becomes $\prod_{k=1}^N\int_0^{\infty}\!d\rho_k\int_0^{2\pi}\!d\varphi_k/2\pi$. The new degrees of freedom $m_k$ are non-negative occupation numbers living on the bond connecting the time slices $k$ and $k+1$. With the (exact) representation of ${\cal Z}_h$ as a product over $1-\Delta h$, i.e.\ before having re-exponentiated $\Delta h$ in Eq.~\eqref{Eq.uru}, we get 
\begin{align}
 {\cal Z}^{(N)}_{h_{q}}
 &=\prod_{k=1}^N
 \sum_{m_k=0}^\infty
 \int_0^{2\pi}\!\!\frac{d\varphi_k}{2\pi}\, 
  {\rm e}^{\,-i\varphi_k(m_k-m_{k-1})}
 \\
 &\times 
 \frac{1}{m_k!}
 \int_0^\infty \!\!\!d\rho_k\,{\rm e}^{-\rho_k}\,
 \rho_k^{(m_k+m_{k-1})/2}
 \big(1-\Delta h_{q}(\rho_k)\big)\,,
\nonumber
\label{Eq.esp}
\end{align} 
where $m_0=m_N$.
Now all original variables $(\rho,\varphi)_k$ can be integrated out exactly: the $\varphi_k$-integrals are that of Lagrange multipliers constraining the $m_k$'s to be identical for neighboring and thus all $k$'s -- which is a manifest conservation of the particle number (as $\hat{H}$ commutes with $\hat{n}$) -- such that only one $m$-sum is left in
\begin{equation}
 {\cal Z}^{(N)}_{h_{q}}
 =\sum_{m=0}^{\infty}\prod_{k=1}^{N}\left[1-\Delta {\cal H}_{q}(m)\right]
 \to \sum_{m=0}^{\infty}{\rm e}^{-\beta {\cal H}_{q}(m)},
\label{Eq.fra}
\end{equation}
where the $\rho_k$-integrals provide the mapping 
\begin{align}
 h(\rho)\:\Rightarrow\:{\cal H}(m)
 =\frac{1}{m!}
 \int_0^\infty\!d\rho\, {\rm e}^{-\rho} \, \rho^m \, h(\rho)\,,
\label{Eq.aus} 
\end{align}
a Laguerre transform with the occupation number as new argument. For $\hat{H}_q$, Eq.~\eqref{Eq.mar}, we get ${\cal H}_{q}(m)=g\, m!/(m-q)!$ making the partition function, Eq.~(\ref{Eq.fra}), {\it exact}. The linearity of Eq.~\eqref{Eq.aus} allows to define ${\cal H}$ symbols for linear combinations of $\hat{H}_q$ and thus all particle-number conserving (one site-)systems. Remarkably, the new symbol ${\cal H}(m)$ has the same functional form as the Hamiltonian $\hat{H}(\hat{n})$. This is a nontrivial result of using 
the $h$-symbol
together with its Laguerre transform to ${\cal H}$ \cite{Supplement}.

With the same formalism we can easily show that, on the contrary, the path integral with a diagonal $H$-symbol, $g\,\rho_k^q$, i.e.\ action (I) in Fig.~\ref{Fig.wrong}, is mapped onto $g\, (m+q)!/m!$ and thus misses ground states, see Fig.~\ref{Fig.comparison} \cite{Supplement}.

The observation that allows us to further {\it reformulate the path integral} is the presence of a term ${\rm e}^{-im_k(\varphi_k-\varphi_{k+1})}$ resembling the polar form of the symplectic term. In order to take full advantage of this observation we now consider a general $L$-mode (or $L$-site) bosonic system described by field operators collected in $\hat{{\bf b}}=(\hat{b}_{1},\ldots,\hat{b}_{L})$ with canonical commutation relations $[\hat{b}_{i},\hat{b}_{j}^{\dagger}]=\delta_{ij}\hat{I}$ and a completely general Hamiltonian $\hat{H}=H(\hat{{\bf b}},\hat{{\bf b}}^{\dagger})$. As for the one-site case, the proper starting point of any exact manipulation is the repeated use of the $P$-representation with symbol $h({\bf z}_{k},{\bf z}_{k}^{*})$ depending on all the site indexes at discrete time $k$, ${\bf z}_{k}=(z_{1,k},\ldots,z_{L,k})$. The construction of ${\cal Z}_{\rm h}^{(N)}$ is straightforward, with a symplectic term $\sum_{k=1}^{N} {\boldsymbol z}_{k}^{*}{\boldsymbol z}_{k+1}$ that is dualized as in Eq.~(\ref{Eq.por}) in terms of non-negative integers ${\boldsymbol m}_{k}$ on time-bonds.  

The key technical step is to transform the sums over discrete dual variables into integrals by means of Poisson resummation \cite{Kleinert,Berry}, i.e., 
\begin{align}
 \sum_{m=0}^{\infty}\! f(m)\,{\rm e}^{-i m(\varphi-\varphi')}
 =\sum_{s=-\infty}^{\infty} 
 \int_{0-0}^\infty\!\!\!\!dm\,
 f(m)\,{\rm e}^{-i m(\varphi-\varphi'+2\pi s)} 
\label{Eq.arg}
 \end{align}
and instead of performing the integrations over both $({\boldsymbol \rho},{\boldsymbol \varphi})_{k}$ (a very difficult task in this general case), we perform exactly only the $\rho$-ones. A major virtue of this exact manipulations is that one can write the integers $s$ as differences of yet other integers, ${\boldsymbol s}_{k}={\boldsymbol r}_{k}-{\boldsymbol r}_{k+1}$ (for $k=2,...,N$), and ${\boldsymbol s}_{1}=-{\boldsymbol r}_{2}$, such that the only appearance of the phases $\varphi$ is next to the corresponding $r$, $({\boldsymbol \varphi}+2\pi {\boldsymbol r})_{k}=:{\boldsymbol \phi}_{k}$ (for $k=2,..,N$). The new angles $\phi$ now extend over the whole axis (eating the $r$-sums), but boundary terms have to be treated separately ${\boldsymbol r}_{N+1}=:{\boldsymbol Q}$, ${\boldsymbol \varphi}_{1}=:{\boldsymbol \varphi}$. Renaming ${\boldsymbol m}_{k}=\boldsymbol{\rho}_{k}$
and shifting for simplicity 
${\boldsymbol \phi}_{k}\to\boldsymbol{\phi}_{k-1}$, and with  
\begin{align}
 {\boldsymbol \phi_0}:={\boldsymbol \varphi},
 \:\:
 {\boldsymbol \phi}_{N}:={\boldsymbol \varphi}+2\pi {\boldsymbol Q},
 \quad
 {\boldsymbol \rho}_{0}:={\boldsymbol \rho}_{N}=:{\boldsymbol \rho}
\label{Eq.ice} 
\end{align}
making winding numbers $Q$ explicit, we get finally
\begin{align}
 {\cal Z}^{(N)}_{{\cal H}}
 &=
 \sum_{{\boldsymbol Q}=-\infty}^\infty  
 \int_0^\infty\!\!\!d{\boldsymbol \rho}
 \int_0^{2\pi}\!\!\!\frac{d{\boldsymbol \varphi}}{(2\pi)^{L}}
 \left[\prod_{k=1}^{N-1}\int_0^\infty\!\!\!\!\!d{\boldsymbol \rho}_k \int_{-\infty}^\infty\!\!\frac{d{\boldsymbol \phi}_k}{(2\pi)^{L}}\right] \nonumber\\
 &\times  
 {\rm e}^{\,
 \sum_{k=0}^{N-1} \left[i
 {\boldsymbol \rho}_{k+1}({\boldsymbol \phi}_{k+1}-{\boldsymbol \phi}_{k})
 -\Delta  {\cal H}\left(
 {\boldsymbol \rho}_{k+1},{\boldsymbol \rho}_{k};
 {\boldsymbol \phi}_{k}\right)\right]
 }\,, 
\label{Eq.per}
\end{align}
(assuming all symbols to be $2\pi$-periodic in $\arg {\boldsymbol z}$) where the new, ``Laguerre'', symbol
\begin{eqnarray}
\label{Eq.den}
&&{\cal H}({\boldsymbol \rho},{\boldsymbol \rho}';{\boldsymbol \phi})= \\ && \left[\prod_{i=1}^{L}
 \int_0^\infty\!dr_{i}\, \frac{{\rm e}^{-r_{i}}r_{i}^{\frac{\rho_{i}+\rho_{i}'}{2}}}{\sqrt{\Gamma(\rho_{i}+1)\Gamma(\rho_{i}'+1})}\right] 
 h({\bf z}^{*},{\bf z}) \Big|_
 {z_{i}=\sqrt{r_{i}}\,{\rm e}^{i\phi_{i}}} \nonumber
\end{eqnarray}
generalizes Eq.~\eqref{Eq.aus}, making \eqref{Eq.per} exact. Thanks to this exactness, the ambiguity of formal manipulations characteristic of the transition to the continuum is now fully resolved by providing a definition of the continuum limit with the discretization prescription of Eq.~\eqref{Eq.per}. Performing the ${\boldsymbol Q}$-sum quantizing $\rho$ to a non-negative integer ${\boldsymbol n}$, this {\it exact} path integral in polar coordinates reads
\begin{align}
 {\cal Z}_{{\cal H}}
 &=\sum_{{\bf n}=0}^{\infty}
 \int_{0}^{2\pi}\!\!\!\frac{d{\boldsymbol \varphi}}{(2\pi)^{L}}
 \int_
 {{\boldsymbol \rho}(\beta-0)={\boldsymbol \rho}(0-0)={\bf n}}
 ^{{\boldsymbol \phi}(\beta)={\boldsymbol \phi}(0)={\boldsymbol \varphi}}
 {\cal D}[{\boldsymbol \rho}(\tau);{\boldsymbol \phi}(\tau)] \nonumber \\ 
 & \times \exp\left(-(S_{\rm s}+S_{\rm d})[{\boldsymbol \rho}(\tau);{\boldsymbol \phi}(\tau)]
 \right)\,,
 \label{Eq.cro}
\end{align}
where we introduced the (in general non-continuous!) $L$-component paths ${\boldsymbol \rho}(\tau), {\boldsymbol \phi}(\tau)$ such that 
${\boldsymbol \rho}(\tau_{k}-\Delta/2):={\boldsymbol \rho}_{k}$, 
${\boldsymbol \phi}(\tau_{k}):={\boldsymbol \phi}_{k}$ 
with $\tau_{k}=k\Delta$ ($k=0,..,N$) and used central difference, midpoint respectively left point rule to obtain
\begin{align}
\label{Eq.nig}
 S_{\rm s}[{\boldsymbol \rho};{\boldsymbol \phi}]
 &:=-\,i\! 
 \int_{0}^{\beta}\!\!d\tau\,
 {\boldsymbol \rho}(\tau)\dot{{\boldsymbol \phi}}(\tau)
 =\,i\! 
 \int_{0}^{\beta}\!\!d\tau\,
 \dot{{\boldsymbol \rho}}(\tau){\boldsymbol \phi}(\tau)\,,\\
 S_{\rm d}[{\boldsymbol \rho};{\boldsymbol \phi}]
 &:= \int_{0}^{\beta}\!d\tau\, 
 {\cal H}\big({\boldsymbol \rho}(\tau+0)
 ,{\boldsymbol \rho}(\tau-0);{\boldsymbol \phi}(\tau)\big)\,,
 \label{Eq.nig2}
\end{align}
including integration by parts (exactly transformed into summation by parts and therefore perfectly valid contrary to the claims of \cite{Kordas}).

{\it Summarizing our main findings so far}, by means of dualization we have i) completely removed any unjustified assumption about path continuity in the coherent state path integral, ii) constructed its exact polar version in both discrete and continuous form and, iii) identified the exact Hamiltonian symbol. 

In particular, for the hopping term of the Bose-Hubbard model, $\hat{H}_{J}=(J/2)\sum_{i,j}(\hat{b}^{\dagger}_{i}\hat{b}_{j}+\hat{b}^{\dagger}_{j}\hat{b}_{i})$ -- where the heuristic symbols suggested by WG do not apply -- the path integral in polar coordinates is exact both in discrete and continuous form if and only if one uses 
\begin{align}
 {\cal H}_{J}({\boldsymbol \rho},{\boldsymbol \rho}';{\boldsymbol \phi})
 &=J\sum_{i,j}\gamma(\rho_{i},\rho_{i}')\gamma(\rho_{j},\rho_{j}')\cos{(\phi_{j}-\phi_{i})}\,,
 \nonumber\\
 \gamma(\rho,\rho')
 &=\frac{\Gamma\left(\frac{\rho+\rho'}{2}+\frac{3}{2}\right)}
 {\sqrt{\Gamma\left(\rho+1\right) \Gamma\left(\rho'+1\right)}}\,.
 \label{Eq.crc}
\end{align}
In the presence of such hopping terms the ${\boldsymbol \phi}$-integrations are that of Lagrange multipliers only after expanding $e^{-S_d}$ (in the coupling $J$) yielding powers of $\int_0^\beta\!d\tau'{\cal H}_{J}(\tau')$. The corresponding factors $e^{\pm i \phi(\tau')}$ give rise to unit jumps in $\rho(\tau)$ (of opposite direction at site $i$ and $j$) at 
those $\tau'$s, a path integral consequence of
$[\hat{H}_{J},\hat{n}_{i}]\ne 0$. The factors 
\begin{align}
\gamma(\rho_i,\rho_i+1)\gamma(\rho_j,\rho_j-1)
=\sqrt{(\rho_i+1)\rho_j}
\label{Eq.mex}
\end{align}
have a characteristic form: the root of the bigger occupation number at each jump, reflecting expectation values of $\hat{b}^{\dagger}\hat{b}$ in number states. While repeated use of this result in the $J$ expansion of ${\cal Z}$ provides an exact resummation in terms of discontinuous paths of the diagrammatic Monte Carlo method \cite{Boninsegni,Prokofev}, the standard replacement $\gamma(\rho,\rho')\simeq \sqrt{\rho}$, only valid for smooth paths and in the semiclassical limit $\rho \gg 1$ \cite{Baranger}, results in expansion coefficients identically equal to zero.  

Our results are of special relevance if amplitude-phase variables are important, a prominent example being {\it spin systems} for which standard derivations of the path integral use the polar (Bloch) representation of the spin coherent states \cite{Schulman,Fradkin, Altland,Kordas2}. As we have seen, such approaches will automatically suffer from the continuity assumption and indeed break down as shown by WG. In the following we will construct the exact path integral for spin systems valid for general Hamiltonians and explicitly show its exactness using dualization.

We consider a spin system described by $\hat{H}_{\text{spin}}$
in terms of operators $\hat{{\bf S}}=(\hat{S}_{x},\hat{S}_{y},\hat{S}_{z})$ obeying $[\hat{S}_{i},\hat{S}_{j}]=i\hbar\epsilon_{ijk}\hat{S}_{k}$ and fixed eigenvalue $\hbar^2 S(S+1)$ of the total spin $\hat{{\bf S}}\cdot \hat{{\bf S}}$. By means of $2\hat{{\bf S}}
=\hbar(\hat{b}_{1}^{\dagger}\hat{b}_{2}+\hat{b}_{2}^{\dagger}\hat{b}_{1},
i(\hat{b}_{1}^{\dagger}\hat{b}_{2}-\hat{b}_{2}^{\dagger}\hat{b}_{1}),\hat{b}_{1}^{\dagger}\hat{b}_{1}-\hat{b}_{2}^{\dagger}\hat{b}_{2})$ 
we represent the spin algebra on an $L=2$ bosonic system, namely on the subspace $\hat{n}_{1}+\hat{n}_{2}=2S$. After we construct the path integral in the whole Hilbert space of the bosonic system, Eqs.~\eqref{Eq.den}-\eqref{Eq.nig}, we identify the different sectors of total particle number to read off the corresponding partition function with fixed total spin. This so-called Schwinger boson mapping avoids the non-analytic symbols from the Holstein-Primakov construction \cite{Auerbach}.

First of all, due to the particular representation of $\hat{{\bf S}}$ in terms of $\hat{{\bf b}}=(\hat{b}_{1},\hat{b}_{2})$ and $\hat{{\bf b}}^{\dagger}$, 
the corresponding symbol $h_{\rm spin}({\bf z},{\bf z}^{*})$ and its Laguerre transform ${\cal H}_{\rm spin}({\boldsymbol \rho},{\boldsymbol \rho}';\phi_{1}-\phi_{2})$ only depend on the angle variable difference. Therefore, we can exactly integrate out the sum $\phi_{1}+\phi_{2}$ that only appears in the symplectic term and that fixes the sum  $\rho_1+\rho_2$ (and thus $n_{1}+n_{2}$) to a constant that is easily identified with $2S$. Writing $\rho_1=\eta$ and $\rho_2=2S-\eta$ (both in $[0,2S]$) and $n_1=n$ (in $\{0,.., 2S\}$), the continuous path integral -- which is 
in one-to-one correspondence with the discrete version -- represents the partition function as 
\begin{align}
 \!\!\!{\cal Z}_{\text{spin}}
 &=\sum_{n=0}^{2S}\int_{0}^{2\pi}\frac{d\varphi}{2\pi} \int_{\eta(\beta-0)=\eta(0-0)=n}^{\phi(\beta)=\phi(0)=\varphi}{\cal D}[\eta\in[0,2S],\phi] \nonumber \\ 
 &\times \exp\left(-S_{\rm s}[\eta;\phi]-S_{\rm d}[\eta,2S-\eta;\phi_{1}-\phi_{2}=\phi]\right)
 \label{Eq.srb}
\end{align}
being the exact path integral for spin systems. Contrary to several claims \cite{Fradkin,Kordas2,Alscher}, no regularization term appears. Moreover, the famous Berry phase/Wess-Zumino-Witten term for spin actions \cite{Berry2,Wess,Witten} follows immediately from $S_s[\eta,\phi]$ upon parametrizing $\eta(\tau)=S(1+\cos\theta(\tau))$.

As a first application, we consider $\hat{H}_{\text{spin}}^{(z)}=f(\hat{S}_{z})=f(\hbar(\hat{n}_1-\hat{n}_2)/2)$ whose symbols $h$ and ${\cal H}$ are independendent of both angular variables, such that $\eta=n$ is constant. Our observation for purely $\hat{n}$-dependent systems  below Eq.~\eqref{Eq.aus} straightforwardly yields ${\cal H}_{\text{spin}}^{(z)}({\boldsymbol \rho})=f(\hbar(\rho_1-\rho_2)/2)=f(\hbar(\eta-S))\equiv f(\hbar(n-S))$, and identifying $n-S$ with the quantum number $m\in\{-S,..,S\}$ gives the exact results for all these cases, whereas WG showed the usual form of the spin path integral to fail except for the special case $S=1/2$ and linear $f$. 

Second, in order to treat another example beyond WG and to see the key importance of the $(\tau\pm 0)$-prescription in the Hamiltonian \eqref{Eq.nig2} (obsolete for the previous case of constant paths), we now consider $S=1/2$ and $\hat{H}_{\text{spin}}^{(x)}=\omega\hat{S}_{x}$ with ${\cal Z}_{\text{spin}}^{(x)}=2\cosh(\beta\hbar\,\omega/2)$. The symbol is that of Eq.~\eqref{Eq.crc} with $J=\hbar\,\omega$ and $\rho_i=\eta\in[0,1]$. To lowest nontrivial order $\omega^2$, two unit jumps of opposite sign (for a periodic $\eta$) appear at $\tau_{1,2}$. The integrand of the $\tau_{1,2}$-integration is the constant $[\hbar\,\omega\,\gamma(0,1)\,\gamma(1,0)/2]^2=(\hbar\,\omega)^2/4$ (see Eq.~\eqref{Eq.mex}) and together with combinatorics this yields the correct perturbative term $(\beta \hbar\,\omega)^2/4$ indeed. Incorrectly using the $H$-symbol forced diagonal again, $H_{\text{spin}}^{(x)}(\eta,\phi)=\hbar\,\omega\sqrt{\eta\,(1-\eta)}\cos{\phi}$, one would have to evaluate $\eta(\tau)$ directly at the jumps $\tau_{1,2}$. Writing $\eta(\tau)$ near a jump with the Heaviside function $\Theta$, the square root becomes $\sqrt{\Theta(0)(1-\Theta(0))}$ and no prescription for $\Theta(0)$ exists in which this square root agrees with $\gamma(0,1)\gamma(1,0)=1$ as necessary for the correct result. 

{\it Outlook}. By constructing the action functionals that render path integrals for general non-relativistic many body bosonic and spin systems exact, we provide the long sought rigorous basis for their systematic use. Two main general aspects of this are i) the numerical implementation of coherent state path integrals based on explicitly summing over paths including regularization of contributions from discontinuities, and ii) semiclassical and saddle-point analysis starting from the correct Hamiltonian symbol. Our construction for spin systems is of particular interest since it can be now be used to construct an exact path integral for {\it fermions} in terms of complex (non-Grassmann) classical functions \cite{Engl}, with the subsequent possibility to construct exact path integrals in condensed matter and also relativistic (supersymmetric, gauge etc.) systems. Extending our dualization methods into {\it real time}, the exact path integral representation of the propagator describing many-body systems far from equilibrium \cite{Tomsovic} can be constructed, which is work in progress.      

\bigskip
\noindent {\it Acknowledgments} -- We thank Quirin Hummel for useful discussions. FB acknowledges support from DFG (Contract
No.\ BR 2872/6-2).

%

\end{document}